\documentclass[preprintnumbers, floatfix, letterpaper, twocolumn,aps,prd,epsfig,nofootinbib,natbib,longbibliography]{revtex4-2}

\usepackage{graphicx}
\usepackage{epstopdf}
\usepackage{natbib}
\usepackage{latexsym}
\usepackage{amssymb}
\usepackage{amsmath}
\usepackage{color}
\usepackage{mathrsfs}
\usepackage{xparse}
\usepackage{bbding}
\usepackage{enumitem}
\usepackage{pifont}
\usepackage[center]{subfigure}
\usepackage[
            pdfstartview=FitH,
            bookmarksnumbered=true,
            bookmarksopen=true,
            colorlinks,
            linkcolor=blue,
            anchorcolor=green,
            citecolor=blue
            ]{hyperref}

            \usepackage{setspace}
\usepackage{amssymb}
\usepackage{graphicx,wrapfig}
\usepackage{color}
\usepackage{mathrsfs}
\usepackage{subfigure}
\usepackage{comment}
\usepackage{multirow}
\usepackage{cleveref}
\usepackage{bm}
\usepackage{soul}
\definecolor{mg}{rgb}{0.0, 0.5, 0.0}

\begin{document}
  \renewcommand\arraystretch{2}
 \newcommand{\bq}{\begin{equation}}
 \newcommand{\eq}{\end{equation}}
 \newcommand{\bqn}{\begin{eqnarray}}
 \newcommand{\eqn}{\end{eqnarray}}
 \newcommand{\nb}{\nonumber}
 \newcommand{\lb}{\label}

\newcommand{\La}{\Lambda}
\newcommand{\va}{\scriptscriptstyle}
\newcommand{\be}{\nopagebreak[3]\begin{equation}}
\newcommand{\ee}{\end{equation}}

\newcommand{\ba}{\nopagebreak[3]\begin{eqnarray}}
\newcommand{\ea}{\end{eqnarray}}

\newcommand{\n}{\nonumber}
\newcommand{\su}{\mathfrak{su}}
\newcommand{\SU}{\mathrm{SU}}
\newcommand{\U}{\mathrm{U}}

\def\be{\nopagebreak[3]\begin{equation}}
\def\ee{\end{equation}}
\def\ba{\nopagebreak[3]\begin{eqnarray}}
\def\ea{\end{eqnarray}}
\newcommand{\f}{\frac}
\def\rmd{\rm d}
\def\pl{\ell_{\rm Pl}}
\def\d{{\rm d}}
\def\fe{\mathring{e}^{\,i}_a}
\def\fw{\mathring{\omega}^{\,a}_i}
\def\fq{\mathring{q}_{ab}}
\def\t{\tilde}

\def\db{\delta_b}
\def\dc{\delta_c}
\def\T{\mathcal{T}}
\def\GammaE{\Gamma_{\rm ext}}
\def\GammaEb{\bar\Gamma_{\rm ext}}
\def\GammaEh{\hat\Gamma_{\rm ext}}
\def\Hee{H_{\rm eff}^{\rm ext}}
\def\H{\mathcal{H}}

\newcommand{\R}{\mathbb{R}}

 \newcommand{\cb}{\color{blue}}
    \newcommand{\cc}{\color{cyan}}
        \newcommand{\cm}{\color{magenta}}
\newcommand{\rc}{\rho^{\scriptscriptstyle{\mathrm{I}}}_c}
\newcommand{\rd}{\rho^{\scriptscriptstyle{\mathrm{II}}}_c}
\NewDocumentCommand{\evalat}{sO{\big}mm}{%
  \IfBooleanTF{#1}
   {\mleft. #3 \mright|_{#4}}
   {#3#2|_{#4}}%
}
\newcommand{\PRL}{Phys. Rev. Lett.}
\newcommand{\PL}{Phys. Lett.}
\newcommand{\PR}{Phys. Rev.}
\newcommand{\CQG}{Class. Quantum Grav.}

\title{Lessons from  gauge fixing and polymerization of loop quantum black holes \\with a cosmological constant}

\author{Geeth Ongole$^{a}$}
\email{geeth\_ongole1@baylor.edu}
\author{Parampreet Singh$^{b}$}
\email{psingh@lsu.edu}
\author{Anzhong Wang$^{a}$}
\email{anzhong\_wang@baylor.edu}

\affiliation{$^{a}$ GCAP-CASPER, Department of Physics and Astronomy, Baylor University, Waco, Texas, 76798-7316, USA\\
$^{b}$ Department of Physics and Astronomy, Louisiana State University, Baton Rouge, Louisiana 70803, USA}

\begin{abstract}
Loop quantization of Schwarzschild black holes with a cosmological constant for 
polymerization parameters which are constant is studied in the effective spacetime description.
We show that for the positive cosmological constant there can be an appearance of large quantum effects at small spacetime curvatures. These effects can manifest as an additional black hole horizon. While the central singularity is resolved in all the cases, these limitations demonstrate incompatibility of the Kantowski-Sachs gauge and schemes with fixed polymerization parameters in the presence of a positive cosmological constant. In contrast, the case of a negative cosmological constant is free of such problematic features. Noted limitations are similar to those in the $\mu_o$ scheme for the loop quantization of cosmological models.

\end{abstract}

\maketitle

\section{Introduction} \label{intro}
\renewcommand{\theequation}{1.\arabic{equation}}
\setcounter{equation}{0}

Quantum geometric effects resulting from loop quantum gravity (LQG) 
have been successfully applied in the context of various cosmological \cite{Ashtekar:2011ni,Li:2023dwy,Agullo:2023rqq} and black hole spacetimes \cite{Gambini:2022hxr,Ashtekar:2023cod}. 
At an operational level, quantum geometry is captured via polymerization of the connection variable and involves  polymerization parameters which are determined from the minimum area of the loop resulting from quantum geometry. The choice of the polymerization parameters appears as a quantization ambiguity in these models. While the non-singular dynamics can be obtained for various choices of polymerization parameters, either constant or phase space dependent, a viable loop quantization must pass through various tests including availability of a consistent quantum gravity scale at which Planck scale effects emerge and compatibility with general relativity (GR) in the infra-red limit \cite{Corichi:2008zb}.  Recent results such as \cite{Li:2021fmu,Motaharfar:2023hil} demonstrate that to test the  viability of a polymerization scheme it is also important to include matter with different values of equation of state.

In the cosmological models, the Friedmann-Robertson-Lemaitre-Walker gauge has been extensively used to study the spacetimes in loop quantum cosmology (LQC). The ambiguity of the choice of the polymerization parameter was settled by demanding certain consistency conditions which ruled out the earlier quantization -- the $\mu_o$ scheme for various reasons, including failure to recover GR in the presence of a positive cosmological constant \cite{Singh:2012zc,Corichi:2008zb}. In the $\mu_o$ scheme the polymerization parameter is constant and in the absence of a positive cosmological constant, not only the singularity is resolved but GR is also recovered at late times. But, the situation changes dramatically when the cosmological constant is included. In the presence of a positive cosmological constant the dynamical equations predict a recollapse instead of an accelerated expansion at late times. The recollapse emerges from fake quantum gravitational effects tied to the polymerization of the connection with a constant parameter. These results  guided the loop quantization towards $\bar \mu$ scheme in which the polymerization parameter depends on square root of the inverse triad. In contrast to the $\mu_o$ scheme, the $\bar \mu$ quantization yields GR for all matter obeying null energy condition and turns 
out to be the unique quantization in LQC  \cite{Corichi:2008zb}, which replaces the big bang singularity by a big bounce \cite{Ashtekar:2006rx,Ashtekar:2006uz,Ashtekar:2006wn,Ashtekar:2006es,Ashtekar:2007em} with a  quantum probability of the bounce which equals unity \cite{Craig:2013mga}.

Extensions of the loop quantization of isotropic models have been studied for anisotropic models, of which one for the Kantowski-Sachs spacetime is of vital interest to explore the quantization of Schwarzschild black holes.  
For this quantization, quantum corrections are tied to two polymerization parameters, namely $\delta_b$ and $\delta_c$. There have been various attempts to loop quantize this spacetime in the symmetry reduced setting \cite{Ashtekar:2005qt, Modesto:2005zm, Boehmer:2007ket, Corichi:2015xia, Olmedo:2017lvt, Ashtekar:2018lag,Ashtekar:2018cay} leading to various explorations in this direction (see for e.g. \cite{Chiou:2008nm,Saini:2016vgo, Yonika:2017qgo, Ashtekar:2020ckv,Bodendorfer:2019xbp,Gan:2020dkb,Giesel:2021dug,Li:2021snn, Garcia-Quismondo:2021xdc,Alonso-Bardaji:2021yls,Ongole:2022rqi,Gan:2022oiy,Zhang:2023noj,Gan:2024rga,Gan:2022mle,Liu:2021djf,Giesel:2024mps,Zhang:2024khj,Ongole:2023pbs}). Of these, the Ashtekar-Olmedo-Singh (AOS) model has various desirable features: independence of physical predictions from fiducial structures, symmetric spacetime across the bounce which translates to identical masses for black hole and white holes, a universal mass independent upper bound on the curvature invariants, and negligible quantum corrections at classical
scales. A study of tidal Love numbers shows that the model is consistent with GR for masses greater than $4.3 \times 10^{4} M_{\mathrm{Pl}}$ \cite{Motaharfar:2025typ}. The AOS model is based on fixing polymerization parameters using Dirac observables approach where the polymerization parameters are functions of the phase space variables in the phase space, but are constant on dynamical trajectories. The goal of this manuscript is to understand the properties of loop quantized Schwarzschild spacetimes in presence of positive and negative cosmological constants which serve as generalization of the AOS model. However, results from the study are applicable to a wider range of models with constant polymerization parameters when connection components are polymerized. The primary focus of our work is to understand whether loop quantized Schwarzschild spacetimes with a positive cosmological constant are free from any undesirable features as is the case for cosmological models in the $\mu_o$ or similar schemes. 

\begin{figure}[htbp]
\includegraphics[width=0.9\linewidth]{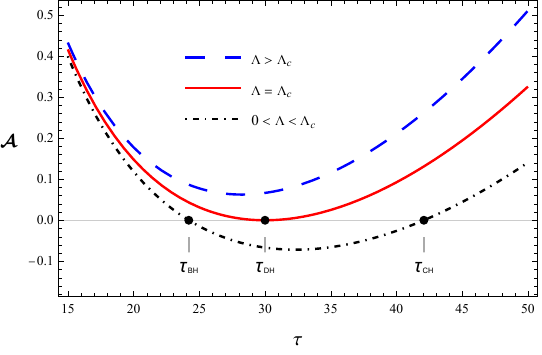}\\
(a)\\
\vspace{.3cm}
 \includegraphics[width=0.9\linewidth]{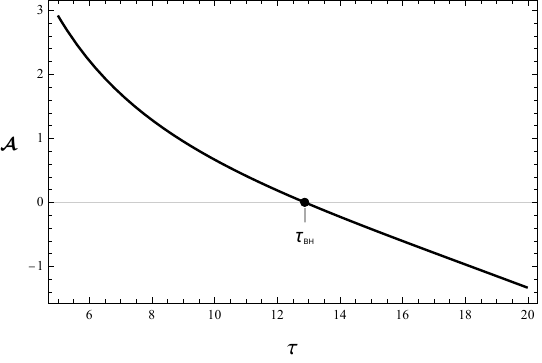}\\
 (b)\\
\caption{(a) Plot of the function ${\cal{A}}(\tau) \left(\equiv \Lambda\tau^2/3   +  2m/\tau - 1\right)$ for $\Lambda > 0$.  The black dot dashed curve represents the $0 < \Lambda < \Lambda_c$ case with two horizons, denoted by $\tau_{\mathrm{BH}}$ and $\tau_{\mathrm{CH}}$ respectively, where $\Lambda_c \equiv 1/(9m^2)$. The red solid curve represents the $\Lambda = \Lambda_c$ case with a degenerate horizon denoted by $\tau_{\mathrm{DH}}$, and the blue dashed curve represents $\Lambda < \Lambda_c$ and has no horizons, so the singularity located at $\tau = 0$ is naked.  (b)  Plot of ${\cal{A}}(\tau)$ for $\Lambda<0$, in which a black hole  horizon always exists. In all the regions where ${\cal{A}}(\tau) > 0$ the spacetime can be written  locally in the Kantowski-Sachs form (\ref{eq2.1}), and the effective loop quantization is applicable to such regions.}
\label{fig1}
\end{figure}

Classically, the spacetimes coupled with a cosmological constant are  the Schwarzschild-de Sitter  solutions for $\Lambda > 0$ and the Schwarzschild-Anti-de Sitter  solutions for $\Lambda < 0$. For the Schwarzschild-de Sitter solutions three different cases raise: $0 < \Lambda < \Lambda_c$, $\Lambda = \Lambda_c$, and $\Lambda > \Lambda_c$,  as shown in Fig. \ref{fig1} (a), where $\Lambda_c \equiv 1/(9m^2)$, with $m$ being the mass parameter of the solutions. In the case  $0 < \Lambda < \Lambda_c$ two horizons exist, being respectively the black hole horizon and the cosmological horizon, denoted by $\tau_{\mathrm{BH}}$ and $\tau_{\mathrm{CH}}$ in Fig. \ref{fig1} (a). In this case, the spacetime is homogeneous in the regions 
$0 < \tau < \tau_{\mathrm{BH}}$ and $\tau > \tau_{\mathrm{CH}}$, and can be cast in the Kantowski-Sachs form (\ref{eq2.1}).  In the case $\Lambda = \Lambda_c$ the two horizon coincide and it will be referred to as the degenerate horizon (DH),  $\tau_{\mathrm{BH}} = \tau_{\mathrm{CH}} \equiv  \tau_{\mathrm{DH}}$. And the surface gravity of such a degenerate horizon is zero. Classically, it is expected that it is not stable against small perturbations. In this case,
the spacetime is homogeneous in the regions $0 < \tau < \tau_{\mathrm{DH}}$ and $\tau > \tau_{\mathrm{DH}}$ and can also be cast in the Kantowski-Sachs form (\ref{eq2.1}).
In the case $\Lambda > \Lambda_c$ the central singularity located at $\tau = 0$ becomes naked, and it is also classically expected not stable. In addition, the entire spacetime  in this case can be cast in the Kantowski-Sachs form (\ref{eq2.1}). On the other hand, in the Schwarzschild-Anti-de Sitter spacetime, a black hole horizon always exists as shown in Fig. \ref{fig1} (b), and the spacetime inside this horizon can also be cast in the Kantowski-Sachs form (\ref{eq2.1}) using the gauge freedom. As a result, loop quantization of black holes can be applied to all the regions in which the corresponding metric can be written in the Kantowski-Sachs form (\ref{eq2.1}). In particular, these include the regions $\tau > \tau_{\mathrm{CH}}$ in the cases $0 < \Lambda \leq \Lambda_c$.

Loop quantization of the Schwarzschild spacetime with a cosmological constant in the Kantowski-Sachs gauge was first studied in \cite{Dadhich:2015ora} using a scheme where polymerization parameters are phase space functions. In a recent work of Ref. \cite{Mato:2024ndc} this spacetime has been studied in the midisuperspace framework  with a Gullstrand–Painlevé-like gauge again using a scheme where the polymerization parameters are functions of the phase space variables. In addition, Ref. \cite{Feng:2024sdo} worked with the combination of the Kantowski-Sachs gauge and constant polymerization parameters, following the AOS model \cite{Ashtekar:2018lag,Ashtekar:2018cay}. 
However, the focus was on fine-tuning a symmetric bounce between the black and white hole horizons.  In addition to this, the quantization outside of the cosmological horizon $(\tau > \tau_{\mathrm{CH}})$ was not considered in that paper. In this manuscript, we systematically study the quantization of all the regions of the Schwarzschild-de Sitter and Schwarzschild-Anti-de Sitter spacetimes that are isometric to the Kantowski-Sachs cosmological model. These include the internal regions $(\tau < \tau_{\mathrm{BH}})$ of black holes,   the external regions of the cosmological horizon $(\tau > \tau_{\mathrm{CH}})$, and the entire spacetime ($\tau > 0$) in the naked singularity case [cf. Fig. \ref{fig1} (a)]. 
To quantize these spacetimes we follow the procedure outlined in \cite{Corichi:2015xia, Olmedo:2017lvt, Ashtekar:2018lag, Ashtekar:2018cay,Ongole:2023pbs} as we choose to work in the Kantowski-Sachs gauge. 
In LQC, effective dynamics plays an important role to extract the resulting physics. It has been shown via extensive numerical studies that effective spacetime description captures the underlying quantum dynamics very accurately \cite{Diener:2014mia,Diener:2017lde}. In this manuscript 
we assume the validity of the effective spacetime description.
We first solve the resulting effective Hamiltonian equations numerically, and then show that the quantized spacetime to the internal of the black hole horizon for macroscopic black holes has negligible quantum corrections near the black hole horizon, and the classical singularity is always replaced by a regular transition surface, which connects smoothly the black hole region to an anti-trapped region, in which a white hole (WH) horizon appears. However, the quantized spacetime outside the cosmological horizon receives considerable quantum corrections and an additional black hole-like horizon appears.  
These are true even when the curvatures of the spacetime are very low and quantum geometric effects are expected to be negligible.  Similar results are also observed in the  critical 
($\Lambda = \Lambda_c$) and naked singularity ($\Lambda > \Lambda_c$) cases.  
A careful study reveals that these undesirable properties are due to the simultaneous adoption of the Kantowski-Sachs gauge when the polymerization parameters are constant on the phase space trajectory. Numerical results that are presented in this paper will use the polymerization parameters as given by \cite{Ashtekar:2018lag, Ashtekar:2018cay}. However, any other choice of polymerization parameters that are constant will give rise to the same results discussed above.

The manuscript is structured as follows. In \Cref{sec2} we write down the classical Hamiltonian of the Schwarzschild spacetime coupled with a cosmological constant. After deriving the classical equations of motion we gauge fix a few integration constants. Then, we find the analytical expressions of the horizon locations for different $\Lambda$ values. In \Cref{sec3} we write the effective Hamiltonian by applying the replacement of Eq.(\ref{eq3.0}) to the classical Hamiltonian given in Eq.(\ref{eq2.3}). The numerical solutions to the effective loop Hamiltonian equations are studied for various cases, beginning with the $\Lambda > \Lambda_c$ case in \Cref{sec3.1}. Later we study the $0 < \Lambda < \Lambda_c$ and $\Lambda = \Lambda_c$ cases in \Cref{sec3.2} and \Cref{sec3.3}, respectively. The $\Lambda <0$ case is discussed in \Cref{sec3.4}. Finally, our main results and some concluding remarks are presented in \Cref{sec4}.

\section{Features of Classical Hamiltonian Dynamics} \label{sec2}
\renewcommand{\theequation}{2.\arabic{equation}}
\setcounter{equation}{0}

The interior of the Schwarzschild spacetime coupled with a cosmological constant  can be written in the Kantowski-Sachs form with the following metric
\begin{align}
    \label{eq2.1}
    \mathrm{d} s^2 = - N_{\tau}^2 \mathrm{d} \tau^2 + \f{p_b^2}{|p_c| L_o^2} \mathrm{d} x^2 + |p_c| \mathrm{d}\Omega^2,
\end{align}
where $N_{\tau},\; p_b$, $p_c$ are only functions of $\tau$ and $\mathrm{d}\Omega^2$ is the metric of the unit 2-sphere with $\mathrm{d}\Omega^2 \equiv \mathrm{d}\theta^2 + \sin^2\theta \mathrm{d}\phi^2$. The Poisson brackets of the phase space variables $p_b$ and $p_c$ are given by
\begin{align}
\label{eq2.2}
    \{c, p_{c}\} = 2G\gamma,  \quad \{b, p_{b}\} = G\gamma,
\end{align}
where $b$ and $c$ are the corresponding momentum conjugates and $G$ is the Newtonian gravitational constant.
The parameter  $\gamma$ denotes the Barbero-Immirzi parameter whose numerical value in this manuscript will be set to $\gamma \approx 0.2375$ following black hole thermodynamics in LQG. 

The corresponding Hamiltonian for the classical Schwarzschild spacetime coupled with a cosmological constant is a function of its phase space variables $\left( b,p_b,c,p_c\right)$ and reads as
\begin{align}
\label{eq2.3}
H_{\mathrm{cl}} &= N_{\mathrm{cl}} \mathcal{H}_{\mathrm{cl}} \nb \\ &= - \frac{N_{\mathrm{cl}}}{2G\gamma^2} \left[2bc\sqrt{p_c}+ \left(b^2 + \gamma^2\right) \frac{p_b}{\sqrt{p_c}} \right] \nb\\
& ~~ + 4\pi N_{\mathrm{cl}} p_b \sqrt{p_c} \; \rho_\Lambda,
\end{align}
where $\rho_\Lambda \equiv \Lambda/8 \pi G$ and $\Lambda$ is the cosmological constant. 

The metric in Eq.(\ref{eq2.1}) is invariant under the following gauge transformation
\begin{align}
    \label{eq2.4}
    \tau = f(\tilde{\tau}), \quad x = \alpha \tilde{x} + x_o,
\end{align}
where $f(\tilde{\tau})$ is an arbitrary function of $\tau$, and $\alpha$ and $x_o$ are constants. With this gauge freedom we choose the following lapse,
\begin{align}
    \label{eq2.5}
    N_{\mathrm{cl}} = \frac{\gamma \mathrm{sgn}(p_c)  |p_c|^{1/2}}{b}.
\end{align}
With the above choice of the lapse, the classical Hamiltonian given by Eq.(\ref{eq2.3}) reads 
\begin{align}
    \label{eq2.6}
    H_{\mathrm{cl}}[N_{\mathrm{cl}}]=-\frac{1}{2G\gamma} \left[2 c p_c +\left(b + \frac{\gamma^2}{b}\right) p_b \right] + \frac{\gamma \Lambda}{2 G}  \frac{p_b p_c}{b}.
\end{align}
Then, the Hamiltonian equations are given by
\begin{align}
    \label{eq2.7}
    \dot b &= G \gamma \frac{\partial H_{\mathrm{cl}}}{\partial p_{b}} = - \frac{1}{2b}(b^2+ \gamma^2) + \frac{\gamma^2 \Lambda}{2}  \frac{p_c}{b},\\
    \label{eq2.8}
    \dot p_{b} &= - G \gamma \frac{\partial H_{\mathrm{cl}}}{\partial b} = \frac{p_{b}}{2b^2} (b^2- \gamma^2) + \frac{\gamma^2 \Lambda}{2} \frac{p_{b}p_{c}}{b^2},\\
    \label{eq2.9}
    \dot c &= 2G \gamma \frac{\partial H_{\mathrm{cl}}}{\partial p_{c}} = - 2c + \gamma^2 \Lambda \frac{p_{b}}{b}, \\
    \label{eq2.10}
    \dot p_{c} &= - 2 G \gamma \frac{\partial H_{\mathrm{cl}}}{\partial c} = 2 p_{c},
\end{align}
where a dot denotes the derivative with respect to $T$, the timelike coordinate  particularly corresponding to the choice of the lapse (\ref{eq2.5}).
To find solutions to the above equations, we solve them in the following order: $p_c, b, p_b$ and using these solutions, we solve for $c$ from the classical Hamiltonian constraint ${\cal H}_{\mathrm{cl}} \approx 0$. The solutions are given by
\begin{align}
\label{eq2.11}
    b(T)^2 &= \gamma^2 f(T), \nb \\
    p_b(T)^2 &= 3 p_{b_{o}}^2 e^{2 T} \gamma^2 f(T),\nb \\
    c(T) &= { \sqrt{3} \gamma^2}p_{b_{o}}\left( \frac{ \Lambda}{3} e^T - \frac{b_1}{2 \gamma^2 p_{c_{o}} e^{2 T}} \right),\nb \\
    p_c(T) &= p_{c_{o}} e^{2 T}, \nb \\
    f(T) &\equiv \frac{ \Lambda}{3}  p_{c_{o}} e^{2T} + \frac{b_1}{\gamma^2 e^{ T}} - 1,
\end{align}
where $p_{b_{o}},\; p_{c_{o}}$ and $b_1$ are integration constants. Without loss of generality we can redefine $p_{b_{o}}$ as $\hat p_{b_{o}} \equiv \sqrt{3}\gamma p_{b_{o}}$. The lapse function in the new $T$ coordinate using Eqs.(\ref{eq2.11}) reads 
\begin{align}
    \label{eq2.12}
    N_{\mathrm{cl}} = \sqrt{\frac{p_{c_o}}{f(T)}} \; e^T,
\end{align}
and the metric in terms of $T$  reads
\begin{align}
    \label{eq2.13}
    \mathrm{d}s^2 =  - \frac{p_{c_o} e^{2T} \mathrm{d}T^2}{f(T)} + \frac{\hat p_{b_o}^2f(T)}{p_{c_o}L_o^2} \mathrm{d}x^2 + p_{c_o} e^{2T}\mathrm{d}^2\Omega.
\end{align}
Setting
\begin{align}
\label{eq2.14}
\tau = \sqrt{p_{c_o}} e^T, \quad {x}  \rightarrow \frac{\gamma \hat p_{b_o}}{\sqrt{p_{c_o}} L_o}x, \quad m \equiv \frac{b_1 \sqrt{p_{c_o}} }{2\gamma^2},
\end{align}
we find that the metric finally takes the familiar form
\begin{align}
\label{eq2.15}
\mathrm{d}s^2 = -  \frac{\mathrm{d}\tau^2}{{\cal{A}}(\tau)} +  {\cal{A}}(\tau) \mathrm{d}x^2 + \tau^2 \mathrm{d}\Omega^2, 
\end{align}
where 
\begin{align}
\label{eq2.16}
 {\cal{A}}(\tau) \equiv \frac{ \Lambda} {3} \tau^2 + \frac{2m}{\tau}  - 1.
\end{align}

In this paper we select the integration constants in Eq.(\ref{eq2.11}) to ensure consistency with the $\Lambda = 0$ case  \cite{Ongole:2023pbs} and they read as 
\begin{align}
    \label{eq2.17}
    c_o &= \gamma L_o/(2r_g), \quad \hat p_{b_o} = - r_g L_o,  \quad p_{c_o} = r_g^2,
\end{align}
where $r_g$ denotes the location of the black hole horizon given by the solution of ${\cal{A}}(\tau) = 0$ \cite{curry1991vacuum}. In the event that a black hole horizon does not exist, we take $r_g$ as a numerical constant whose choice does not affect the qualitative results. The classical solutions in Eq.(\ref{eq2.11}) after fixing the integration constants are given by
\begin{align}
    \label{eq2.18}
    c(T) &=  - \gamma r_g L_o\left( \frac{ \Lambda}{3} e^T - \frac{m}{r_g^3 e^{2 T}} \right),\nb \\
    p_c(T) &= r_g^2 e^{2 T},\nb \\
    b(T) &= \gamma  {\cal{A}}^{1/2}(T),\nb \\
    p_b(T) &=   - r_g L_o e^{T}  {\cal{A}}^{1/2}(T),\nb\\
    {\cal{A}}(T) &=  \frac{ \Lambda}{3}  r_g^2 e^{2T} + \frac{2m}{r_g e^{ T}} - 1.
\end{align}

Before analyzing the solutions,  it is important to find the horizons of the Schwarzschild-de Sitter black hole. To this purpose, we first note that  ${\cal{A}}'(\tau) = 2\Lambda(\tau^3 - 3m/\Lambda)/(3\tau^2)$. Thus, the function ${\cal{A}}(\tau)$ takes its minimal value  at $\tau_{\mathrm{min}} = (3m/\Lambda)^{1/3}$, given by
\begin{align}
    \label{eq2.0018}
    {\cal{A}}_{\mathrm{min}} \equiv  {\cal{A}}(\tau_{\mathrm{min}}) = 3m\left\{\left(\frac{\Lambda}{3m}\right)^{1/3} - \frac{1}{3m}\right\},
\end{align}
for $\Lambda > 0$. For $\Lambda < 0$, we have $\tau_{\mathrm{min}} < 0$, which locates outside of the physical range $\tau \ge 0$. As a result, ${\cal{A}}(\tau)$ has no minimum in the region $\tau \geq 0$, as shown in \Cref{fig1} (b) for $\Lambda < 0$. We divide this into two cases, positive $\Lambda$ and negative $\Lambda$.

{\bf Case $\Lambda > 0$:} In this case, from \Cref{fig1} (a),  we can see that in the region $\tau \geq 0$, there may exist two real positive roots, depending on the values of ${\cal{A}}_{\mathrm{min}}$. In particular, if ${\cal{A}}_{\mathrm{min}} < 0$, there exist two different positive roots, say,  $\tau_{\mathrm{BH}}$ and $\tau_{\mathrm{CH}}$. Without loss of generality, we assume that $\tau_{\mathrm{CH}} > \tau_{\mathrm{BH}}$. Then, $\tau_{\mathrm{CH}}$ is often referred to as the cosmological horizon, while $\tau_{\mathrm{BH}}$  the black hole horizon. Hence, from the definition of $r_g$, we find $r_g = \tau_{\mathrm{BH}}$. Note that in both of the regions $0 < \tau < \tau_{\mathrm{BH}}$ and $\tau > \tau_{\mathrm{CH}}$, the function ${\cal{A}}(\tau)$ is positive, that is, the corresponding solutions can be cast in the Kantowski-Sachs form. The two positive roots become equal when
${\cal{A}}_{\mathrm{min}} = 0$. In this cases, 
${\cal{A}}(\tau)$ is always positive for $\tau (0, \infty)$, except for the point $\tau = \tau_{\mathrm{BH}}$, at which we have ${\cal{A}}(\tau_{\mathrm{BH}}) = 0$, and a degenerate horizon appears.
Since now the surface gravity at the black hole horizon vanishes, $\kappa_+ \equiv {\cal{A}}'(\tau_{\mathrm{BH}})/2 = 0$, then the horizon is expected to be unstable. If ${\cal{A}}_{\mathrm{min}} > 0$, no positive real roots exists, and we have ${\cal{A}}(\tau) > 0$
for any given  $\tau$ in the range $\tau \in (0, \infty)$. Then,
the singularity at $\tau = 0$ becomes naked. Therefore, we have the following cases:
\begin{itemize}
    \item {$0 < \Lambda < \Lambda_c$}: In this case, there are two horizons located at
        \begin{align}
        \label{eq2.19}
        \tau_{\mathrm{BH}} &= \frac{2}{\sqrt{\Lambda}} \cos{\left( \frac{1}{3} \cos^{-1} \left(\sqrt{\frac{\Lambda}{\Lambda_c}}\right) + \frac{\pi}{3}\right)},\nb\\
        \tau_{\mathrm{CH}} &= \frac{2}{\sqrt{\Lambda}} \cos{\left( \frac{1}{3} \cos^{-1} \left(\sqrt{\frac{\Lambda}{\Lambda_c}}\right) - \frac{\pi}{3}\right)}, 
        \end{align}
    where $\Lambda_c \equiv {1}/(9m^2)$. In the limit $m \sqrt{\Lambda} \ll 1$, we have
    \begin{align}
    \label{eq2.0019}
        \tau_{\mathrm{BH}} &\simeq 2 m + \mathcal{O} \left( m^2 \Lambda \right), \nb \\
        \tau_{\mathrm{CH}} &\simeq \sqrt{\frac{3}{\Lambda}}  + \mathcal{O} \left( m \sqrt{\Lambda} \right). 
    \end{align}
    \item {$\Lambda = \Lambda_c$}: In this case, the two horizons coincide and we have a degenerate horizon at $\tau_{\mathrm{BH}} = \tau_{\mathrm{CH}}  = 3m \equiv \tau_{\mathrm{DH}}$.
    \item {$\Lambda > \Lambda_c$}: Now, there are no horizons and the singularity located at $\tau = 0$ is naked. As there no horizons for this case we must choose $r_g$ as some constant. In this paper, we choose $r_g=2m$ when working with this case.
\end{itemize}

{\bf Case $\Lambda < 0$:} In this case, we have 
\begin{align}
\label{eq2.0119}
    {\cal{A}} (\tau) \equiv -\frac{|\Lambda|}{3} \tau^2 + \frac{2 m}{\tau} - 1, 
\end{align}
for which we have ${\cal{A}}'(\tau) = -2|\Lambda|(\tau^3 + 3m/|\Lambda|)/(3\tau^2)$. Thus, now
${\cal{A}}(\tau)$ has a maximum  at $\tau_{\mathrm{max}} = - (3m/|\Lambda|)^{1/3}$ with its maximal value
\begin{align}
    \label{eq2.0219}
    {\cal{A}}_{\mathrm{max}} \equiv  {\cal{A}}(\tau_{\mathrm{max}}) = - 3m\left\{\left(\frac{|\Lambda|}{3m}\right)^{1/3} + \frac{1}{3m}\right\},
\end{align}
as shown in \Cref{fig1} (b). From this figure it can be seen that ${\cal{A}}(\tau)$ always has a positive real root at
\begin{align}
\label{eq2.0319}
    \tau_{\mathrm{BH}} &= \left(\frac{3m}{|\Lambda|} + \sqrt{\left(\frac{1}{|\Lambda|}\right)^{3} + \left(\frac{3m}{|\Lambda|}\right)^2}\right)^{\frac{1}{3}} + \nb \\ 
    & \left(\frac{3m}{|\Lambda|} - \sqrt{\left(\frac{1}{|\Lambda|}\right)^{3} + \left(\frac{3m}{|\Lambda|}\right)^2}\right)^{\frac{1}{3}}.
\end{align}
In the limit $m \sqrt{\Lambda} \ll 1$, we have
\begin{align}
    \label{eq2.0320}
    \tau_{\mathrm{BH}} \simeq 2 m + \mathcal{O} \left( m^2 \Lambda \right).
\end{align}

\section{Features of Effective Quantum Hamiltonian Dynamics} \label{sec3}
\renewcommand{\theequation}{3.\arabic{equation}}
\setcounter{equation}{0}

The effective Hamiltonian that captures the leading-order quantum corrections is obtained by the following polymerization on the classical variables $b$ and $c$,
\begin{align}
\label{eq3.0}
    b \rightarrow \frac{\sin\left(\delta_b b\right)}{\delta_b}, \quad
    c \rightarrow \frac{\sin\left(\delta_c c\right)}{\delta_c},
\end{align}
where $\delta_b$ and $\delta_c$ are the two polymerization parameters. In this paper we shall treat them like constants. With the above replacements in the classical Hamiltonian, the effective Hamiltonian reads 
\begin{align}
    \label{eq3.01}
    H_{\mathrm{eff}} &\equiv N \mathcal{H} \nb \\ &= - \frac{1}{2 G \gamma} \left[2 \frac{\sin (\delta_c c)}{\delta_c} \, p_c + \left(\frac{\sin (\delta_b b)}{\delta_b} + \frac{\gamma^2 \delta_b}{\sin(\delta_b b)} \right) \, p_b \right] \nb \\ &+ \frac{\gamma \Lambda}{2 G} \frac{\delta_{b}p_b p_c}{\sin(\delta_{b}b)},
\end{align}
with the following choice of the lapse function,
\begin{align}
    \label{eq3.02}
    N = \frac{\gamma \delta_b\; {\mathrm{sgn}}\left(p_c\right) \sqrt{|p_c|}}{\sin{\left(\delta_b b\right)}}.
\end{align}
The effective Hamiltonian equations of $H_{\mathrm{eff}}$ are given by,
\begin{align}
\label{eq3.07}
\dot b &= G \gamma \frac{\partial H_{\mathrm{eff}}}{\partial p_{b}}  = - \frac{1}{2} \left( \frac{\sin(\delta_b b)}{\delta_b} + \frac{\gamma^2 \delta_b}{\sin(\delta_b b)} \right)\nb\\
& ~~~~~~~~~~~~~~~~~~~~  
    + \frac{\gamma^2 \Lambda}{2} \frac{\delta_b p_c}{\sin(\delta_b b)}, \\
     \label{eq3.08}
\dot p_b &= - G \gamma \frac{\partial H_{\mathrm{eff}}}{\partial b}   = \frac{p_b}{2} \cos(\delta_b b) \left( 1 - \frac{\gamma^2 \delta_b^2}{\sin^2(\delta_b b)} \right) \nb\\
    &~~~~~
    + \frac{\gamma^2 \Lambda}{2} \frac{\delta_b^2 p_b p_c \cos(\delta_b b)}{\sin^2(\delta_b b)}, \\
     \label{eq3.09}
\dot c &= 2G \gamma \frac{\partial H_{\mathrm{eff}}}{\partial p_{c}} = - 2 \frac{\sin(\delta_c c)}{\delta_c} 
    + \gamma^2 \Lambda \frac{\delta_b p_b}{\sin(\delta_b b)}, \\
     \label{eq3.010}
\dot p_c &= - 2 G \gamma \frac{\partial H_{\mathrm{eff}}}{\partial c} 
= 2 p_c \cos(\delta_c c).
\end{align}
Since there are no known closed form analytical solutions available to solve this system of equations, we will solve them numerically.
Here we outline the process of choosing initial conditions for doing so. These initial conditions are to be carefully chosen so that the constraint ${\cal H}_{\mathrm{eff}} \approx 0$ is always satisfied. 
We choose the initial conditions of three phase space variables  
at the initial time $T_i$ where there is a good agreement of the effective dynamics with GR
\begin{align}
\label{eq3.1}
    \left( p_b(T_i),c(T_i),p_c(T_i) \right)_{\mathrm{eff}} = \left( p_b(T_i),c(T_i),p_c(T_i) \right)_{\mathrm{GR}}.
\end{align}
Using the above three initial conditions, we solve the Hamiltonian constraint ${\cal H}_{\mathrm{eff}} \approx 0$ for the fourth phase space variable $b(T_i)_{\mathrm{eff}}$. At this $T_i$ we ensure that 
\begin{align}
    \label{eq3.009}
    |b(T_i)^{\mathrm{eff}} - b(T_i)^{\mathrm{GR}} | \ll 0, \nb \\
    \delta_b b(T_i) \ll 0,\quad \delta_c c(T_i) \ll 0.
\end{align}
These conditions are usually not mutually exclusive and rather go hand in hand. Further, $T_i$ should not be infinitesimally close to the horizon (if one is present) as this will lead to inconsistent numerical solutions \cite{Gan:2022oiy}. 
In this paper, to ensure the validity of our numerical solutions, we plot the effective Hamiltonian and verify that it is vanishing to a great accuracy. 
We further monitor the deviations between the effective and classical phase space variables at $T_i$ and maintain it below $10^{-6}$ unless otherwise mentioned. It is important to note that even though the deviations at $T_i$ are negligible this does not guarantee effective theory will closely follow classical trajectories at all times as quantum geometry effects become dominant in some regions. Though it is natural to expect quantum geometry effects to be dominant at high curvatures (which is observed in all the cases discussed here), we find that there always exist cases in which  the deviations are significant even in the low curvature regions. These include the regions $\tau_{\text{CH}}$ in the case $0 < \Lambda \leq \Lambda_c$ and the region $\tau \gg 0$ in the case $\Lambda > \Lambda_c$.  This is because the effective theory is dependent on two extra parameters, $\delta_b, \;\delta_c$. These parameters appear as trigonometric functions of $\delta_b b, \delta_c c$ and because of this they exhibit oscillatory behavior. Then, from Eq.(\ref{eq3.02}) we can see that $N^2 \rightarrow \infty$ as $\sin(\delta_b b) = 0$, whereby a horizon 
is developed.  Such oscillations  are absent in the classical theory, and as a result lead the effective theory to deviate from classical one.


\begin{figure*}[htbp]

\includegraphics[width=0.95\linewidth]{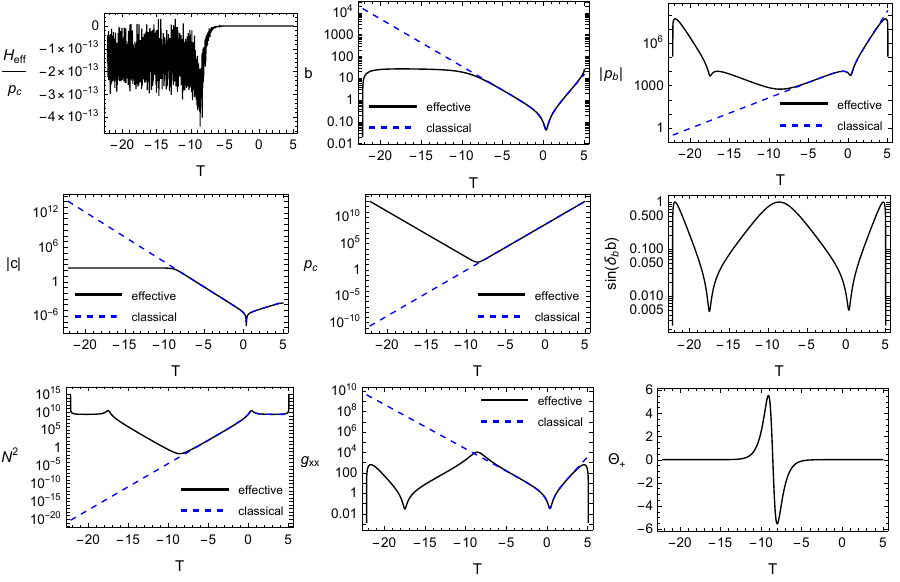}
\caption{The effective dynamics of the phase space variables and metric components for $m=10^4$, $\delta_b=0.1172$, $\delta_c=0.0126$ and $\Lambda/\Lambda_c = 1.1$ for   the $\Lambda > \Lambda_c$ case, in which there appears classically a naked singularity at the center $p_c^{\mathrm{GR}} = 0$. The normalized Hamiltonian is plotted to monitor the accuracy of our numerical results. Plots of $\sin{\left(\delta_b b\right)}$ and $\Theta_+$ are also presented to understand the existence of horizons and transition surfaces. The transition surface is located at $T_{\cal{T}} \simeq -8.5499$, while the white hole and black hole are located at $T_{\mathrm{WH}} \simeq -22.1844$ and $T_{\mathrm{BH}} \simeq 5.0845$, respectively. The initial conditions used to produce these plots are set at $T_i = 0.5$.}
\label{Fig4}
\end{figure*}

\begin{figure*}[htbp]
\includegraphics[width=0.95\linewidth]{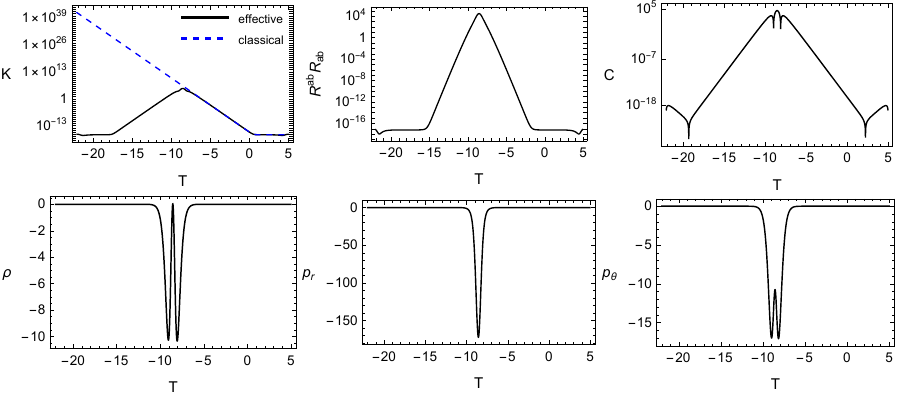}
\caption{Curvature invariants such as the Kretschmann scalar $K (\equiv R_{\alpha\beta\gamma\delta}R^{\alpha\beta\gamma\delta}$), $R^{ab} R_{ab}$, the Weyl scalar $C (\equiv C_{\alpha\beta\gamma\delta}C^{\alpha\beta\gamma\delta})$, the energy density $\rho$, and the radial and tangential pressures $p_r$ and $p_{\theta}$ of the effective energy-momentum tensor for $m=10^4$, $\delta_b=0.1172$, $\delta_c=0.0126$ and $\Lambda/\Lambda_c = 1.1$. }
\label{Fig4.5}
\end{figure*}

\begin{figure}[!htbp]
    \centering
    \includegraphics[height=21.5cm]{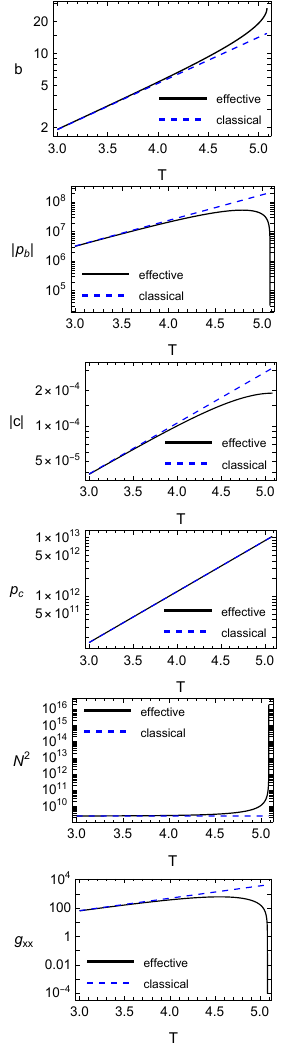}
    \caption{Plots of the phase space variables $b,p_b, c$ and  the metric components $N^2, g_{xx}$ show deviations from the classical trajectories near the black hole horizon located at $T_{\mathrm{BH}} \simeq 5.0845$ for $m=10^4$, $\delta_b=0.1172$, $\delta_c=0.0126$ and $\Lambda/\Lambda_c = 1.1$.}
    \label{newfig4}
\end{figure}

 
\subsection{\texorpdfstring{$\Lambda > \Lambda_c$}{Lambda > Lambda\_c}} \label{sec3.1}

In this case, we have ${\cal{A}}(\tau) > 0$, as shown in Fig. \ref{fig1} (a).
Therefore, the classical singularity located at $\tau = 0$ is naked. To understand the quantum effects to this naked singularity, let us first consider  the representative case $m=10^4$ and $\Lambda / \Lambda_c = 1.1$.
Since there are no horizons present in this  case, one can choose the initial conditions at any time $T_i$, as long as the conditions outlined in Eq.(\ref{eq3.009}) are satisfied. We have also observed that choosing the initial conditions at a point either smaller or greater than the minima of ${\cal{A}} (\tau)$ does not change the spacetime structure.

We plot the Hamiltonian in Fig. \ref{Fig4} to monitor the accuracy of our numerical calculations. The normalized Hamiltonian ($H_{\mathrm{eff}}/p_c$)  is of the order $10^{-13}$ validating our results. Further, the erstwhile naked singularity is now replaced by a smooth and regular  transition  surface, located at $T_{\cal{T}} \simeq -8.5499$, at which we have $\dot{p}_c = 0$, as seen in the plot of $p_c$ in Fig. \ref{Fig4}. To understand the nature of this bounce, let us first calculate the expansions of the ingoing and outgoing null rays \cite{Baumgarte:2010ndz,Hawking:1973uf,Ashtekar:2018cay,Wang:2003bt,Wang:2003xa}, given by 
\begin{align}
\label{eq3.11old}
    \Theta_{\pm} = - \frac{\dot p_{c}}{\sqrt{2} N p_c} =- \frac{\dot p_{c} \sin{\left(\delta_b b\right)}}{\sqrt{2} \gamma \delta_b  p_c^{3/2}}.
\end{align}
Now it is clear from the above expression that $\Theta_{\pm}$ will be zero either when $\dot p_{c}=0$ or $N \rightarrow \infty$. At the transition  surface $p_c$ does attain its minima and one can see from the plot of $\Theta_{+}$ in Fig. \ref{Fig4} that it does indeed go to zero. Since the phase space variables, the metric components and the scalar invariants are all finite at $T_{\cal{T}}$, as shown by Figs. \ref{Fig4} and \ref{Fig4.5}, we conclude that the singularity is replaced by a regular transition surface. Note that  the energy density $\rho$ and radial and tangential pressures $p_r$ and $p_{\theta}$ are defined from  the effective energy-momentum tensor $\kappa T^{\mathrm{eff}}_{\mu\nu} \equiv G_{\mu\nu}$, with
\bqn
\lb{eq3.11new}
\kappa T^{\mathrm{eff}}_{\mu\nu} = \rho u_{\mu} u_{\nu} + p_r x_{\mu}x_{\nu} + p_{\theta}\left(\theta_{\mu}\theta_{\nu} +\phi_{\mu}\phi_{\nu}\right), ~~~~
\eqn
where ($u_{\mu}, x_{\mu}, \theta_{\mu},\phi_{\mu}$) are the unit vectors along respectively the ($T, x, \theta, \phi$)-directions.

Now, moving away from the transition  surface toward the increasing direction of $T$, that is, for $T > T_{\cal{T}}$, there is a great overlap between effective and classical phase space variables. However, near the point $T_{\mathrm{BH}} \simeq 5.0845$, the metric components $N^2$ and $g_{xx}$ and the phase variables $b, c, p_b$ deviate from their classical values dramatically, as can be seen from Fig. \ref{newfig4}. 
In particular, we find that $N^2 \rightarrow \infty$, $\sin(\delta_b b) \rightarrow 0$ and $\Theta^{\pm} \rightarrow 0$, while all the scalar invariants remain finite [cf. Fig. \ref{Fig4.5}]. This indicates a black hole-like horizon is formed at $T_{\mathrm{BH}}$. This is the case no matter how large $m$ is and how far $T_i$ from $T_{\cal{T}}$. Although physically it is hard to understand this, mathematically it can be understood as follows: For 
$T \gg T_{\cal{T}}$ it is expected that $b(T) \simeq b(T)_{\mathrm{GR}} \propto e^{T}$, as shown in Fig. \ref{Fig4}. Then, as $T$ continuously increases, there will exist a moment $T_{\mathrm{BH}}$ such that $\delta_b b(T_{\mathrm{BH}}) = \pi$, at which the lapse function $N (\propto 1/\sin(\delta_b b)$ diverges, whereby a black hole horizon is formed. 

On the other hand, in the region $T < T_{\cal{T}}$, $b(T)$ remains almost constant for $ T \gtrsim T_{\mathrm{WH}}\; (\simeq -22.1844)$, but suddenly drops to zero at $T \simeq T_{\mathrm{WH}}$. Note that as $T \rightarrow T_{\mathrm{WH}}$, we have $N^2(T) \rightarrow \infty$ and $\Theta_{+} \rightarrow 0$ again, while all the scalar invariants remains finite, as shown in Fig. \ref{Fig4}. Hence, a white hole-like horizon is formed.

In addition to the above representative case, we have also studied other cases and observed the similar properties of the corresponding spacetimes.  
In particular, in \Cref{Table4} and \Cref{Table5} we present the locations of the transition surfaces and the black hole and white hole horizons with different choices of $m$ and $\Lambda$ for $\Lambda > \Lambda_c$.
It should be noted that in Figs. \ref{Fig4} - \ref{newfig4} and \Cref{Table4} and \Cref{Table5} we considered only the cases with $(\delta_b, \delta_c) = (0.1172, 0.0126)$. But, for other choices we found similar behaviors.

\begin{table}[htbp]
\centering
\begin{tabular}{|c|c|c|c|}
\hline
\(\frac{\Lambda}{\Lambda_c}\) & \(T_{\text{WH}}\) & \(T_{\cal{T}}\) & \(T_{\text{BH}}\) \\
\hline
1.1 & -22.1844 & -8.5499 & 5.0845 \\
1.5 & -22.0293 & -8.5499 & 4.9294 \\
2 & -21.8855 & -8.5499 & 4.7856 \\
10* & -21.0806 & -8.5499 & 3.9809 \\
100* & -19.9289 & -8.5496 & 2.8296 \\
1000* & -18.7768 & -8.5493 & 1.6783 \\
\hline
\end{tabular}
\caption{Table for $m=10^4$, $\delta_b=0.1172$, $\delta_c=0.0126$ for
$\Lambda > \Lambda_c$, in which a classical naked singularity always exists.
But, after taking the quantum geometric effects into account, it is 
always replaced by a regular transition surface located at $T_{\cal{T}}$.  The table also gives the locations of the white hole horizon ($T_{\text{WH}}$) and black hole  ($T_{\text{BH}}$)
horizons. The asterisk (*) on the $\Lambda / \Lambda_c$ values indicate considerable deviations between the classical and effective phase space variables at the initial time $T_i$. }
\label{Table4}
\end{table}

\begin{table}[htbp]
\centering
\begin{tabular}{|c|c|c|c|}
\hline
\(\frac{\Lambda}{\Lambda_c}\) & \(T_{\text{WH}}\) & \(T_{\cal{T}}\) & \(T_{\text{BH}}\) \\
\hline
1.1 & -29.8596 & -11.6201 & 6.6194 \\
1.5 & -29.7045 & -11.6201 & 6.4644 \\
2 & -29.5607 & -11.6201 & 6.3205 \\
10 & -28.7559 & -11.6201 & 5.5158 \\
100 & -27.6046 & -11.6201 & 4.3645 \\
1000* & -26.4533 & -11.6200 & 3.2132 \\
\hline
\end{tabular}
\caption{Table for $m=10^6$, $\delta_b=0.0252$, $\delta_c=0.0027$ for
$\Lambda > \Lambda_c$, in which a classical naked singularity always exists.
But, after taking the quantum geometric effects into account, it is 
always replaced by a regular transition surface located at $T_{\cal{T}}$.  The table also gives the locations of the white hole horizon ($T_{\text{WH}}$) and black hole  ($T_{\text{BH}}$)
horizons. The asterisk (*) on the $\Lambda / \Lambda_c$ values indicate considerable deviations between the classical and effective phase space variables at the initial time $T_i$.}
\label{Table5}
\end{table}

\subsection{\texorpdfstring{$0 < \Lambda < \Lambda_c$}{0 < Lambda < Lambda\_c}} \label{sec3.2}

In the case $0 < \Lambda < \Lambda_c$, the classical spacetime has two horizons: the black hole horizon and the cosmological horizon, denoted respectively by $\tau_{\mathrm{BH}}$ and $\tau_{\mathrm{CH}}$ in Fig. \ref{fig1} (a). From the figure we can see that ${\cal{A}}(\tau) > 0$ in the two regions $0 < \tau < \tau_{\mathrm{BH}}$ and $\tau > \tau_{\mathrm{CH}}$.  As a result, the corresponding metric in these two regions can be always cast in the Kantowski-Sachs form (\ref{eq2.1}). Then, the quantization presented above can be applied to any of these two regions. A natural question is how the spacetime looks like after the quantization in each of the two regions? To answer this question, we shall numerically solve Eqs.(\ref{eq3.07}) - (\ref{eq3.010}) in each of these two regions, and then study the main properties of the resulting spacetime.   Given the distinct spacetime structure of these two regions, in the following let us study them separately.

\begin{figure*}[htbp]
\includegraphics[width=0.95\linewidth]{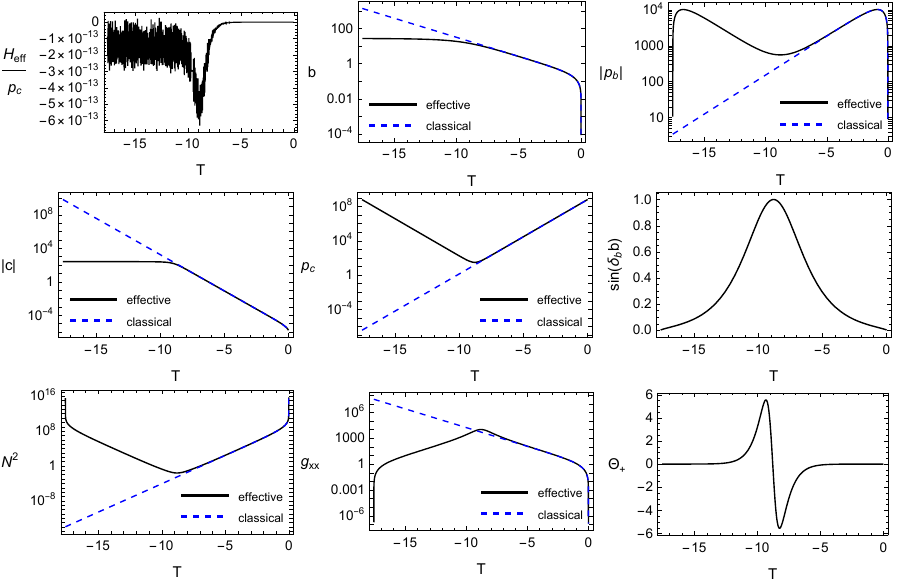}
\caption{The dynamics of the phase space variables in the black hole interior ($T < T_{\mathrm{BH}}$) for $m=10^4$, $\delta_b=0.1172$, $\delta_c=0.0126$ and $\Lambda/\Lambda_c = 0.9$. The normalized Hamiltonian is plotted to monitor the accuracy of our numerical results, and additional plots are for the lapse function $N^2$, the metric component  $g_{xx}$ and the quantities  $\sin{\left(\delta_b b\right)}$  and $\Theta_+$.  The initial conditions used to produce these plots is set at $T_i = -10^{-3}$, where the classical black hole horizon is located at $T_{\mathrm{BH}} = 0$.
 A transition surface, a black hole and a white hole horizon are developed at
$T_{\cal{T}} \simeq -8.79595$,  $T_{\mathrm{BH}} \simeq 0$  and $T_{\mathrm{WH}} \simeq -17.579$, respectively.}
\label{Fig6}
\end{figure*}

\begin{figure*}[htbp]
\includegraphics[width=0.95\linewidth]{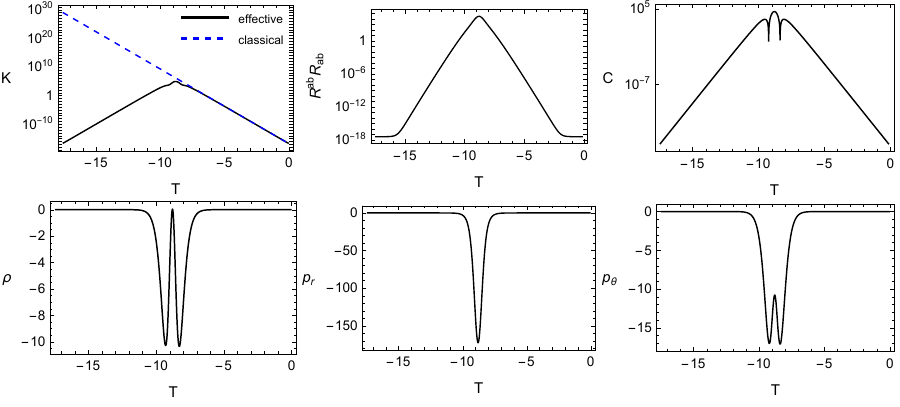}
\caption{Curvature invariants such as the Kretschmann scalar ($K$), $R^{ab} R_{ab}$, the Weyl scalar ($C$) and the radial and tangential pressures $p_r$ and $p_{\theta}$ of the effective energy-momentum tensor in the black hole interior ($T < T_{\mathrm{BH}}$) for $m=10^4$, $\delta_b=0.1172$, $\delta_c=0.0126$ and $\Lambda/\Lambda_c = 0.9$.}
\label{Fig6.5}
\end{figure*}

\subsubsection{Spacetime in the Region $ T <  T_{\mathrm{BH}}$}

Classically, the black hole horizon is given by Eq.(\ref{eq2.19}). Using the gauge freedom (\ref{eq2.4}), without loss of the generality, we can always set $T_{\mathrm{BH}}(\tau_{\mathrm{BH}}) = 0$,  a condition that we shall adopt in this subsection. We numerically solve the system of Eqs.(\ref{eq3.07}) - (\ref{eq3.010}) for various choices of the mass parameter $m$ and the cosmological constant $\Lambda$ to understand the consequences of the quantization in the region $T < T_{\mathrm{BH}}$. We present the results for $m=10^4,\,\Lambda / \Lambda_c = 0.9$  
in Figs. \ref{Fig6} and \ref{Fig6.5}. Inline with earlier discussions, we plot the normalized Hamiltonian to monitor the errors in our numerical results. Dynamics of the phase space variables suggest the central black hole singularity is resolved with a transition surface, as seen in the plots of $p_c$ and $\Theta_+$ in Fig. \ref{Fig6}, as $p_c$ reaches a minimum and all other phase space variables remain finite. The quantum corrections in the form of $\sin{\left(\delta_b b\right)}$ shown in Fig. \ref{Fig6}, are maximum at bounce as one expects in this region of high curvature.

There is also a great overlap between the effective and classical trajectories in the trapped region, the right-hand side of the transition surface,  $T > T_{\cal{T}}$. However, $\Theta_+$ approaches zero at $T \simeq 0$,  
confirming the presence of a black hole horizon even after the quantum geometric effects are taken into account. One can alternatively look at the plot of $\sin{\left(\delta_b b\right)}$ to identify the location of the black hole horizon by determining when it approaches to zero. As expected, this is not the only horizon, as $\sin{\left(\delta_b b\right)}$ and hence $\Theta_+$ approach zero again in the left-hand side of the transition surface, indicating the presence of an anti-trapped or white hole horizon. We also present the plots of a few important curvature invariants such as Kretschmann scalar, $R^{ab} R_{ab}$, the Weyl scalar  $C$, and the energy density $\rho$ and the radial and tangential pressures $p_r$ and $p_{\theta}$ of the effective energy-momentum tensor in Fig. \ref{Fig6.5}. These properties are shared with various choices of $m$ and $ \Lambda$. In particular, we present the results for a few other choices in \Cref{Table6,Table7} where we have the locations of the white hole horizon, transition surface and the black hole horizon.

\begin{table}[htbp]
\centering
\begin{tabular}{|c|c|c|c|}
\hline
\(\frac{\Lambda}{\Lambda_c}\) & \(T_{\mathrm{WH}}\) & \(T_{\cal{T}}\) & \(T_{\mathrm{BH}}\) \\
\hline
0.9 & -17.579 & -8.79595 & 0 \\
0.09 & -17.1215 & -8.56376 & 0 \\
0.009 & -17.0967 & -8.5513 & 0 \\
9 \(\times\) 10\(^{-7}\) & -17.0941 & -8.54996 & 0 \\
9 \(\times\) 10\(^{-12}\) & -17.0941 & -8.54996 & 0 \\
\hline
\end{tabular}
\caption{Table for $m=10^4$, $\delta_b=0.1172$, $\delta_c=0.0126$ for the case $0 < \Lambda < \Lambda_c$. The table gives the locations of the WH horizon ($T_{\mathrm{WH}}$), the transition surface ($T_{\cal{T}}$) and the black hole horizon ($T_{\mathrm{BH}}$) for various choices of $\Lambda$, where $\Lambda_c\equiv \frac{1}{9m^2}$ is the critical value of $\Lambda$.}
\label{Table6}
\end{table}

\begin{table}[htbp]
\centering
\begin{tabular}{|c|c|c|c|}
\hline
\(\frac{\Lambda}{\Lambda_c}\) & \(T_{\text{WH}}\) & \(T_{\tau}\) & \(T_{\text{BH}}\) \\
\hline
0.9 & -23.7313 & -11.8661 & 0 \\
0.09 & -23.2674 & -11.6339 & 0 \\
0.009 & -23.2424 & -11.6214 & 0 \\
9 \(\times\) 10\(^{-8}\) & -23.2398 & -11.6201 & 0 \\
9 \(\times\) 10\(^{-11}\) & -23.2398 & -11.6201 & 0 \\
\hline
\end{tabular}
\caption{Table for $m=10^6$, $\delta_b=0.0252406$, $\delta_c=0.0027187$ for the case $0 < \Lambda < \Lambda_c$. The table gives the locations of the white hole horizon ($T_{\mathrm{WH}}$), the transition surface ($T_{\cal{T}}$) and the BH horizon ($T_{\mathrm{BH}}$) for various choices of $\Lambda$, where $\Lambda_c\equiv \frac{1}{9m^2}$ is the critical value of $\Lambda$.}
\label{Table7}
\end{table}

\begin{figure*}[htbp]
\includegraphics[width=0.95\linewidth]{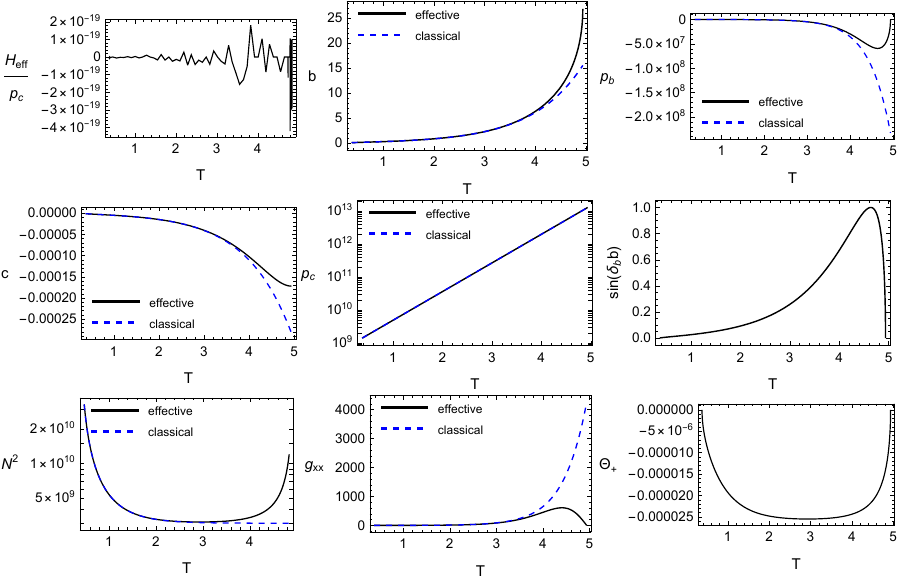}
\caption{The dynamics of the phase space variables in the region $T > T_{\mathrm{CH}}$ for $m=10^4$, $\delta_b=0.1172$, $\delta_c=0.0126$ and $\Lambda/\Lambda_c = 0.9$. The normalized Hamiltonian is plotted to monitor the accuracy of our numerical results, and additional plots are for the lapse function $N$, the metric component  $g_{xx}$, the quantities  $\sin{\left(\delta_b b\right)}$ and $\Theta_+$.  Deviation between classical and effective trajectories can be seen clearly in the later stages of evolution ($T \gtrsim 3$). The initial condition used to produce these plots is set at $T_i = 0.6$, though the spacetime is independent of this choice. Location of the cosmological horizon is at $T_{\mathrm{CH}} \simeq 0.3774$, while a new black hole-like horizon appears at $T_{\mathrm{BH}} \simeq 4.9388$.}
\label{Fig8}
\end{figure*}

\begin{figure*}[htbp]
\includegraphics[width=0.95\linewidth]{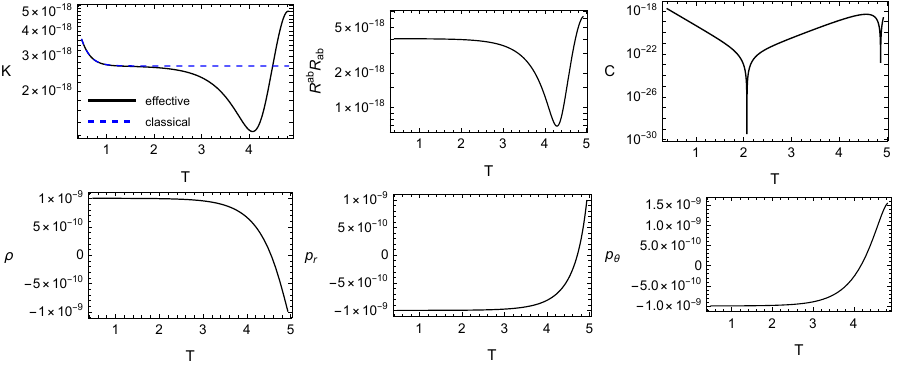}
\caption{Curvature invariants such as the Kretschmann scalar ($K$), $R^{ab} R_{ab}$, the Weyl scalar ($C$) and the radial and tangential pressures $p_r$ and $p_{\theta}$ of the effective energy-momentum tensor in the region $T > T_{\mathrm{CH}}$ for $m=10^4$, $\delta_b=0.1172$, $\delta_c=0.0126$ and $\Lambda/\Lambda_c = 0.9$.}
\label{Fig8.5}
\end{figure*}

\subsubsection{Spacetime in the Region  $T > T_{\mathrm{CH}}$} 

In this subsection, we study quantum geometric effects on the spacetime in the region $T > T_{\mathrm{CH}}$. 
Numerical studies of the spacetime with a choice of $T_i$, where $T_i > T_{\mathrm{CH}}$ are carried out and the results for $m=10^4,\, \Lambda / \Lambda_c = 0.9$   are shown in Figs. \ref{Fig8} and \ref{Fig8.5}. From Fig. \ref{Fig8}  we can see the dynamics of the phase-space variables to the right-hand side of the cosmological horizon show a clear deviation between the classical and effective trajectories for the later stages of the evolution. Further, from Fig. \ref{Fig8} one can see $\sin{\left(\delta_b b\right)}$ and $\Theta_+$ approach zero near the cosmological horizon confirming the existence of a cosmological horizon as predicted from the classical theory. However, the quantum geometric effects lead to the formation of another horizon at $T_{\mathrm{BH}} > T_{\mathrm{CH}}$,  as  both $\Theta_+$  and $\sin{\left(\delta_b b\right)}$ go to zero, where $T_{\mathrm{BH}} = 4.9388$. The reason for this to happen is precisely as that pointed out in the naked singularity case, that is, for $T \gg T_{\mathrm{CH}}$  we have $b(T) \simeq b(T)_{\mathrm{GR}} \propto e^{T}$. Then, there always exists a moment $T_{\mathrm{BH}}$, so that $\delta_b b = \pi$ for any given constant $\delta_b$. Then, we have  $\sin{\left(\delta_b b\right)} \simeq 0$, $\Theta^{\pm} = 0$, and $N^2 \rightarrow \infty$. On the other hand,  from Fig. \ref{Fig8.5}  we can see that all the curvature invariants remain finite, indicating a black hole-like horizon is formed. 

Such a horizon is always formed no matter how low the curvatures of the spacetime are. 
This is not physically expected, and a careful investigation shows that this is due to both the choice of the Kantowski-Sachs gauge and the use of  schemes where the polymerization parameters are either constant on the entire phase space or on the dynamical trajectories \cite{Ashtekar:2018lag,Ashtekar:2018cay}. As a matter of fact, because of the Kantowski-Sachs gauge, the phase variable $b(T)_{\mathrm{GR}} \propto e^T$ [cf. Eq.(\ref{eq2.11})] is increasing monotonically. In the low curvature region, we expect $b(T) \simeq b(T)_{\mathrm{GR}}$. Then,
there always exists a moment $T_{\mathrm{BH}} \;(> T_{\mathrm{CH}})$, such that 
\bq
\lb{eq3.11}
\delta_b b = \pi,  \quad \Theta^{\pm} = 0, \quad N^2 \rightarrow \infty.
\eq
This is true no matter the classical singularity is naked or not, as long as it is in a region far away from the central singularity. Clearly, this is not physical and must be discarded. On the other hand, 
if $\delta_b$ and $\delta_c$ are functions of the phase variables, these undesirable features might be avoided, for example, the hybrid scheme, considered recently in \cite{Gan:2024rga}.  

It should be noted that the $\bar\mu$-scheme often used in LQC is also not compatible with the Kantowski-Sachs gauge, as they always lead to large quantum geometric effects near the classical black hole horizon, so that   finally  the region $T > T_{\cal{T}}$ becomes geodesically complete \cite{Saini:2016vgo} and a black hole horizon does not exist any more, after the quantum geometric effects are taken into account \cite{Zhang:2023noj,Gan:2022oiy}, no matter how low the curvature will be at the classical black hole horizon.

\subsection{\texorpdfstring{$\Lambda = \Lambda_c$}{Lambda = Lambda\_c}} \label{sec3.3}

In this subsection we study the  case $\Lambda = \Lambda_c$ where the classical spacetime has a degenerate horizon, $\tau_{\mathrm{BH}} = \tau_{\mathrm{CH}}$, located at $\tau_{\mathrm{DH}}\; (\equiv \tau_{\mathrm{BH}})$ as shown in Fig. \ref{fig1} (a). Classically, this horizon has zero surface gravity, and is expected not stable. However, quantum mechanically we find that it still exists even after the quantum geometric effects are taken into account.
To show this, let us first note that this horizon naturally divides the whole spacetime into two regions, $T > T_{\mathrm{DH}}$ and $T < T_{\mathrm{DH}}$, in which  ${\cal{A}} (T) > 0$ is always true. Therefore, the spacetime in each of them can be cast in the Kantowski-Sachs form (\ref{eq2.1}). As a result, the quantization presented above is applicable to each of these two regions. Similar to the last case, let us study the quantization in each of them  separately.

\begin{figure*}[htbp]
\includegraphics[width=0.95\linewidth]{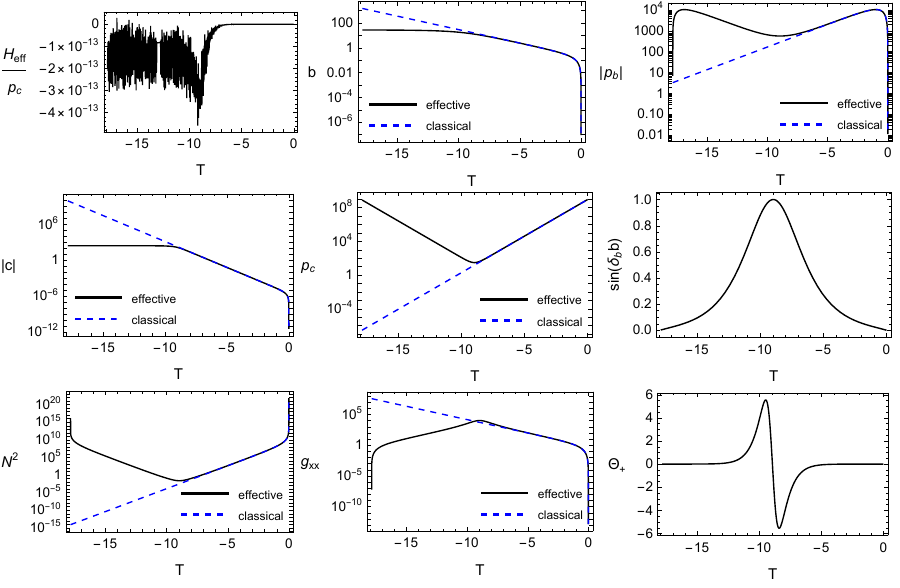}
\caption{The dynamics of the phase space variables in the region $T < T_{\mathrm{DH}}$ for  $m=10^4$, $\delta_b=0.1172$, $\delta_c=0.0126$ and $\Lambda = \Lambda_c$. The normalized Hamiltonian is plotted to monitor the accuracy of our numerical results, and additional plots are for the lapse function $N$, the metric component  $g_{xx}$, the quantities  $\sin{\left(\delta_b b\right)}$ and $\Theta_+$. The initial condition used to produce these plots is set at $T_i = -10^{-4}$. The central singularity now is replaced by a regular transition surface located at $T_{\cal{T}} \simeq -8.9554$. The location of a black hole-like  horizon is at $T_{\mathrm{BH}} \simeq 0$, while
the location of a white hole-like  horizon is at $T_{\mathrm{WH}} \simeq -17.846$.}
\label{Fig10}
\end{figure*}

\begin{figure*}[htbp]
\includegraphics[width=0.95\linewidth]{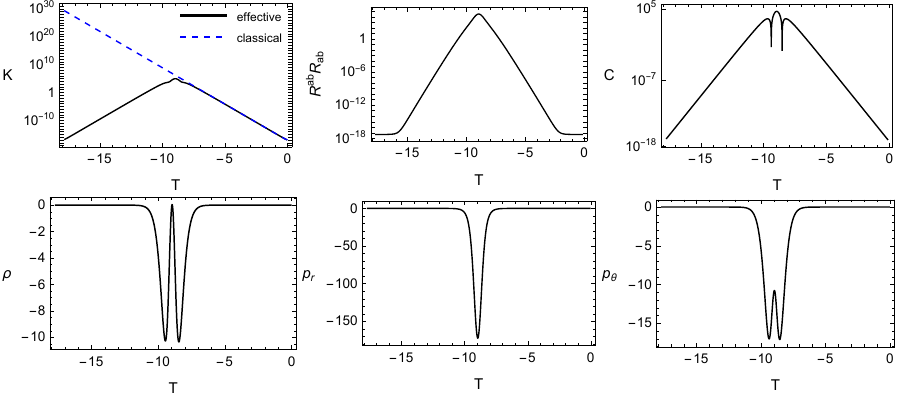}
\caption{Curvature invariants such as the Kretschmann scalar ($K$), $R^{ab} R_{ab}$, the Weyl scalar ($C$) and the radial and tangential pressures $p_r$ and $p_{\theta}$ of the effective energy-momentum tensor in the region $T < T_{\mathrm{CH}}$  for   $m=10^4$, $\delta_b=0.1172$, $\delta_c=0.0126$, in the case $\Lambda = \Lambda_c$.}
\label{Fig10.5}
\end{figure*}

\subsubsection{Spacetime in the Region $T < T_{\mathrm{DH}}$}

For comparison with the cases studied earlier, we also choose $m=10^4$  in this case and  present the results in Figs. \ref{Fig10} and \ref{Fig10.5}. From these figures, we can see that the properties of the corresponding spacetime are very similar to the  case $\Lambda < \Lambda_c$ in the region inside the black hole horizon.  In particular, the central singularity is replaced by a regular transition surface, located  at $T_{\cal{T}} = -8.9554$. A black hole-like horizon still exists, located at $T_{\mathrm{BH}} \simeq 0$, while in the region $T < T_{\cal{T}}$ a WH is developed at $T_{\mathrm{WH}} \simeq -17.846$.

\subsubsection{Spacetime in the Region $T > T_{\mathrm{DH}}$} 

\begin{figure*}[htbp]
\includegraphics[width=0.95\linewidth]{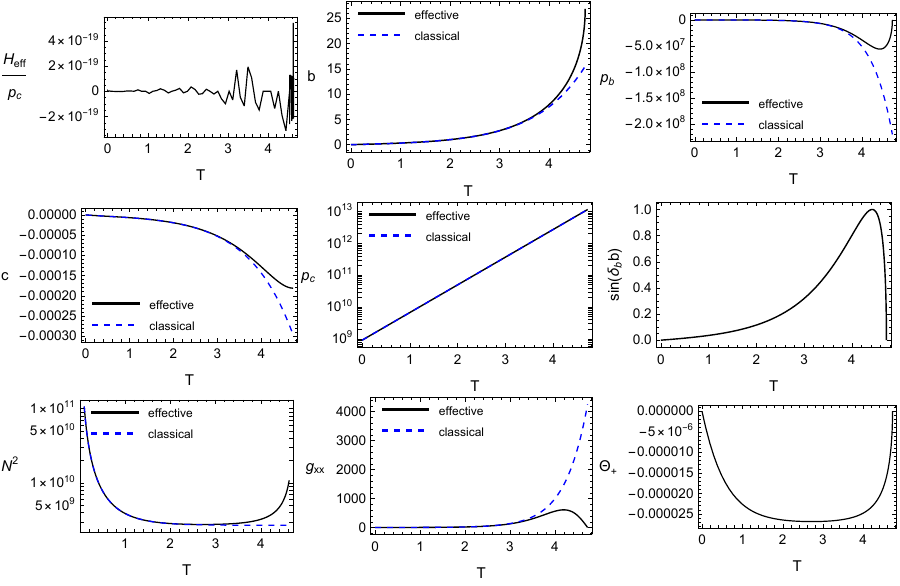}
\caption{The dynamics of the phase space variables in the region $T > T_{\mathrm{DH}}$ for  $m=10^4$, $\delta_b=0.1172$, $\delta_c=0.0126$ and $\Lambda = \Lambda_c$. The normalized Hamiltonian is plotted to monitor the accuracy of our numerical results, and additional plots are for the lapse function $N$, the metric component  $g_{xx}$, the quantities  $\sin{\left(\delta_b b\right)}$ and $\Theta_+$. The initial condition used to produce these plots is set at $T_i = -10^{-4}$.  Location of the degenerate horizon is at $T_{\mathrm{DH}} = 0$, while a new black hole-like horizon is located at  
$T_{\mathrm{BH}} \simeq 4.7267$.}
\label{Fig12}
\end{figure*}

\begin{figure*}[htbp]
\includegraphics[width=0.95\linewidth]{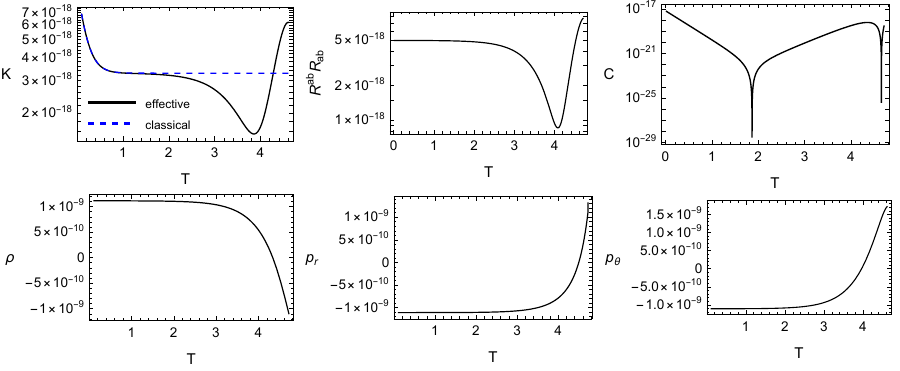}
\caption{Curvature invariants such as the Kretschmann scalar ($K$), $R^{ab} R_{ab}$, the Weyl scalar ($C$), and the energy density $\rho$, the radial and tangential pressures $p_r$ and $p_{\theta}$ of the effective energy-momentum tensor in the region $T > T_{\mathrm{CH}}$  for  $m=10^4$, $\delta_b=0.1172$, $\delta_c=0.0126$ and $\Lambda = \Lambda_c$.}
\label{Fig12.5}
\end{figure*}

In this region,  we show the physical quantities  for  $m=10^4$ in Figs. \ref{Fig12} and \ref{Fig12.5}. From these figures we can see that the properties are similar in nature to the $T > T_{\mathrm{CH}}$ in the $\Lambda < \Lambda_c$ case. In particular, in addition to the presence of a degenerate horizon, a black hole-like horizon is always developed at $T_{\mathrm{BH}} = 4.7267$, at which Eq.(\ref{eq3.11}) becomes true. With the same reasons, we contribute this to the simultaneous use of the Kantowski-Sachs gauge and  schemes where the polymerization parameters are either constant on the entire phase space or on the dynamical trajectories.

\subsection{\texorpdfstring{$\Lambda < 0$}{Lambda < 0}} \label{sec3.4}

\begin{figure*}[htbp]
\includegraphics[width=0.95\linewidth]{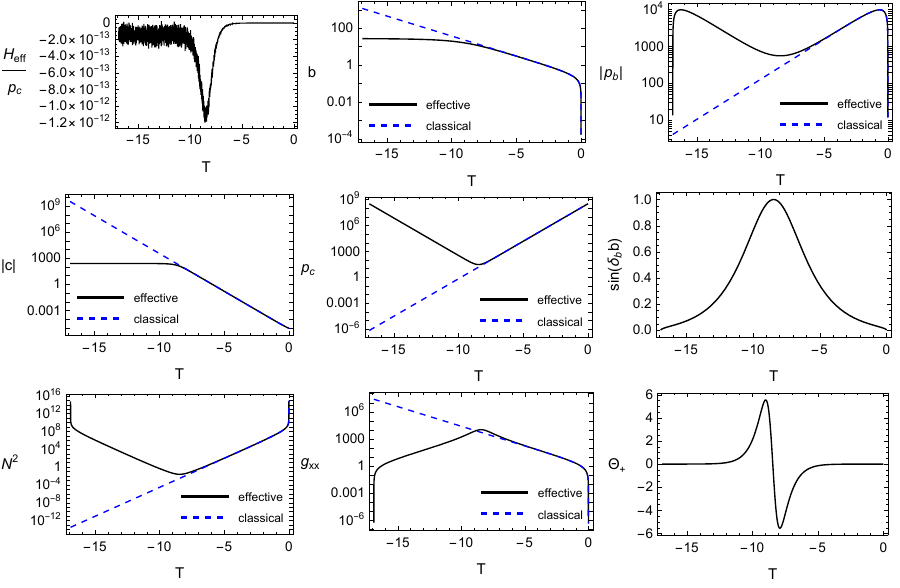}
\caption{The dynamics of the phase space variables in the region $T < T_{\mathrm{BH}}$ for  $m=10^4$, $\delta_b=0.1172$, $\delta_c=0.0126$ and $\Lambda = -10^{-9} < 0$. The normalized Hamiltonian is plotted to monitor the accuracy of our numerical results, and additional plots are for the lapse function $N$, the metric component  $g_{xx}$, the quantities  $\sin{\left(\delta_b b\right)}$ and $\Theta_+$. The initial condition used to produce these plots is set at $T_i = -10^{-12}$.  Location of the black hole   is at $T_{\mathrm{BH}} = 0$, while the transition surface and the white hole horizon  are located respectively at  $T_{\cal{T}} \simeq -8.4469$ and $T_{\mathrm{WH}} \simeq -16.889$.}
\label{Fig14}
\end{figure*}

\begin{figure*}[htbp]
\includegraphics[width=0.95\linewidth]{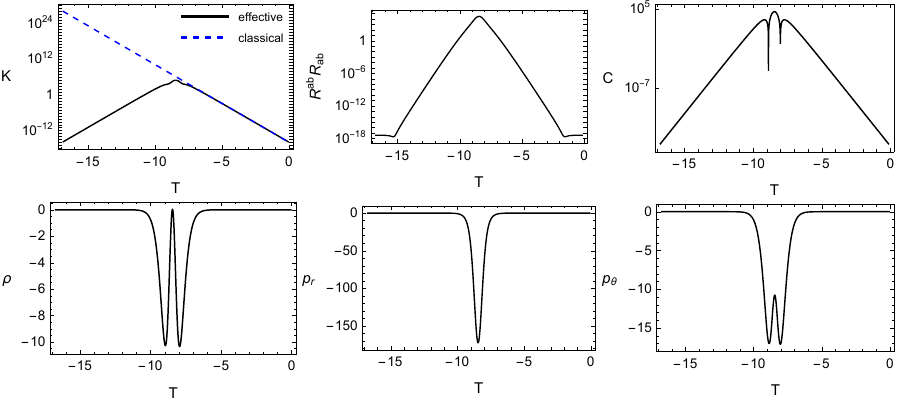}
\caption{Curvature invariants such as the Kretschmann scalar ($K$), $R^{ab} R_{ab}$, the Weyl scalar ($C$), and the energy density $\rho$ and radial and tangential pressures $p_r$ and $p_{\theta}$ of the effective energy-momentum tensor in the region $T < T_{\mathrm{BH}}$  for  $m=10^4$, $\delta_b=0.1172$, $\delta_c=0.0126$ and $\Lambda = -10^{-9} < 0$.}
\label{Fig14.5}
\end{figure*}

Here we consider the quantization of the anti-de Sitter Schwarzschild  spacetime and present our numerical results for $m=10^4, \Lambda = -10^{-9} $ in Figs. \ref{Fig14} and \ref{Fig14.5}. Since classically we have only one horizon, the quantization considered above is only applicable to the region inside this horizon. Then, we find that the general behavior of the interior spacetime is similar in nature to the black hole interior of the $\Lambda < \Lambda_c$ de Sitter Schwarzschild spacetime. The quantum geometric effects resolve the central singularity and replace it by a regular transition surface. A good overlap between classical and effective trajectories is observed in the right-hand region  ($T > T_{\cal{T}} \simeq -8.4469$)  of the transition surface in Fig. \ref{Fig14}. A look at the plot of $\Theta_+$  shows the presence of two horizons, a black hole horizon at $T_{\mathrm{BH}} \simeq 0$, as expected from the classical theory and a white hole horizon located at $T_{\mathrm{WH}} \simeq -16.889$ in the left-hand side of the transition surface. From the plot of $\sin{\left(\delta_b b\right)}$ in Fig. \ref{Fig14}, one can see it goes to zero at both of these points which results in the lapse function diverging and whereby confirming the presence of black hole/white hole horizons. A good overlap on the right-hand side  of the transition surface is also confirmed from the plot of the Kretschmann scalar  as shown in Fig. \ref{Fig14.5}. This finiteness is also evident in the plots of $R^{ab} R_{ab}$ and the Weyl scalar. Additionally, we also plot the components of the energy momentum tensor in Fig. \ref{Fig14.5}.

In \Cref{Table8} and \Cref{Table9} we present  the locations of the white hole horizon, transition surface and the black hole horizon for $m=10^4$ and $m=10^6$ respectively for a few choices of $\Lambda$.

\begin{table}[htbp]
\centering
\begin{tabular}{|c|c|c|c|}
\hline
\(\frac{\Lambda}{\Lambda_c}\) & \(T_{\mathrm{WH}}\) & \(T_{\cal{T}}\) & \(T_{\mathrm{BH}}\) \\
\hline
\(-1\) & -4.60566 & -2.38417 & 0 \\
\(-10^{-1}\) & -6.14495 & -3.10984 & 0 \\
\(-10^{-2}\) & -7.68307 & -3.85903 & 0 \\
\(-10^{-3}\) & -9.21476 & -4.61579 & 0 \\
\(-10^{-5}\) & -12.2389 & -6.12189 & 0 \\
\(-10^{-10}\) & -17.0684 & -8.53705 & 0 \\
\(-10^{-15}\) & -17.0941 & -8.54996 & 0 \\
\(-10^{-20}\) & -17.0941 & -8.54996 & 0 \\
\hline
\end{tabular}
\caption{Table for $m=10^4$, $\delta_b=0.1172$, $\delta_c=0.0126$ for $\Lambda < 0$.  Locations of a white hole horizon ($T_{\mathrm{WH}}$), transition surface ($T_{\cal{T}}$) and a BH horizon ($T_{\mathrm{BH}}$) for various choices of $\Lambda < 0$ are presented. The black hole horizon is always set to $T_{\mathrm{BH}} = 0$ using the gauge freedom.}
\label{Table8}
\end{table}

\begin{table}[htbp]
\centering
\begin{tabular}{|c|c|c|c|}
\hline
\(\frac{\Lambda}{\Lambda_c}\) & \(T_{\mathrm{WH}}\) & \(T_{\cal{T}}\) & \(T_{\mathrm{BH}}\) \\
\hline
\(-1\) & -4.60566 & -2.38495 & 0 \\
\(-10^{-1}\) & -6.14842 & -3.11132 & 0 \\
\(-10^{-3}\) & -9.22842 & -4.62216 & 0 \\
\(-10^{-5}\) & -12.2992 & -6.15135 & 0 \\
\(-10^{-10}\) & -19.8434 & -9.92177 & 0 \\
\(-10^{-15}\) & -23.2371 & -11.6187 & 0 \\
\(-10^{-20}\) & -23.2398 & -11.6201 & 0 \\
\(-10^{-22}\) & -23.2398 & -11.6201 & 0 \\
\hline
\end{tabular}
\caption{Table for $m=10^6$, $\delta_b=0.0252406$, $\delta_c=0.0027187$ for $\Lambda < 0$.  Locations of a white hole horizon ($T_{\mathrm{WH}}$), transition surface ($T_{\cal{T}}$) and a black hole horizon ($T_{\mathrm{BH}}$) for various choices of $\Lambda$ are presented. The black hole horizon is always set to $T_{\mathrm{BH}} = 0$ using the gauge freedom.}
\label{Table9}
\end{table}

\section{Conclusions} \label{sec4}
\renewcommand{\theequation}{4.\arabic{equation}}
\setcounter{equation}{0}

Loop quantization of cosmological models where the connection variable is polymerized using a constant polymerization parameter is known to be problematic in the presence of a positive cosmological constant \cite{Singh:2012zc,Corichi:2008zb}. While the classical big bang singularity is resolved and replaced by a big bounce, the growth in connection at late times causes a recollapse of the universe in striking contrast to the classical behavior where the universe dominated by a positive cosmological constant would expand forever. Such an unphysical behavior demonstrated 
limitations of the $\mu_o$ scheme in LQC and led to the improved dynamics or the $\bar \mu$ scheme \cite{Ashtekar:2006wn,Ashtekar:2007em}. Using these schemes gravitational collapse scenarios have been investigated. While the $\bar \mu$ scheme yields singularity resolution along with various novel features and are consistent with GR at small spacetime curvature scales (see for eg. \cite{Giesel:2021dug}), $\mu_o$ scheme remains problematic. In particular, formation of trapped surfaces is ruled out for the homogenous dust collapse in marginally bound case  \cite{Li:2021snn}.

In this manuscript we investigated the viability of schemes for the loop quantization of Schwarzschild spacetime in presence of a cosmological constant where the connection components are polymerized with constant polymerization parameters. Our analysis used effective spacetime description in the 
Kantowski-Sachs gauge, for which the metric
can be cast in the form (\ref{eq2.1}). The properties of the effective Hamiltonian crucially depend on the choices of the polymerization parameters $\delta_b$ and $\delta_c$, and in this manuscript we fixed these parameters using AOS approach which has various notable features in the absence of cosmological constant, including symmetric bounce. Note that in order to get a symmetric bounce, i.e. same mass for black holes and white holes, one needs to fine tune $\delta_b$ and $\delta_c$ in presence of the cosmological constant and strictly speaking our choice of parameters do not give an exactly symmetric bounce. Nevertheless, our conclusions are valid for any choices of these parameters as long as  they are constant.

Using these constant polymerization parameters,  we have systematically studied the dynamical equations (\ref{eq3.07}) - (\ref{eq3.010}) in each of the cases, $\Lambda > \Lambda_c \; [\equiv 1/(9m^2)]$, $0 < \Lambda < \Lambda_c $, $\Lambda = \Lambda_c$ and $\Lambda < 0$, and found that, whenever a black hole horizon exists classically (this, for example, is the case for $\Lambda \leq \Lambda_c$),  the quantized spacetime inside the black hole horizon  is quite similar to the case without the cosmological constant considered in \cite{Ashtekar:2020ifw,Ongole:2023pbs}, in which the classical singularity is always replaced by a regular transition surface that connects two regions, one is trapped and the other is anti-trapped. Within a finite distance from the transition surface a black hole horizon exists in the trapped region, while a white hole exists in the anti-trapped region. However, to our great surprise, a new black hole-like horizon always exists in the region far away from the central singularity, no matter how small the curvature of the spacetime becomes.  In the case $0 < \Lambda < \Lambda_c$,  it is in the region $T > T_{\mathrm{CH}}$, and in the case $\Lambda = \Lambda_c$,  it is in the region $T > T_{\mathrm{DH}}$. Such a horizon also exists in the naked singularity case $\Lambda > \Lambda_c$, in the region $T \gg 0$. To understand this, let us consider the region far away from the central singularity. In such a region, the curvature of the spacetime is very low, and we would expect that the quantum geometric effects are negligible, so we have 
  \bq
  \lb{eq4.1}
  b(T) \simeq b(T)_{\mathrm{GR}} = \gamma {\cal{A}}^{1/2}(T) \simeq \sqrt{\frac{\gamma^2  r_g^2 \Lambda}{3}}  e^{T},
  \eq
 for $\Lambda > 0$ and  $T \gg T_i$. Assuming that $\delta_b b(T_i) > 0$, then we find that there always exists a moment $T_{\mathrm{BH}} > T_i$, such that
 \bq
 \lb{eq4.2}
 \delta_b b(T_{\mathrm{BH}})  \simeq  \delta_b \sqrt{\frac{\gamma^2  r_g^2\Lambda}{3}}  e^{T_{\mathrm{BH}}} = \pi, 
 \eq
 at which we have 
 \bq
\lb{eq4.3}
 N^2(T_{\mathrm{BH}}) \rightarrow \infty,  \quad \Theta^{\pm}(T_{\mathrm{BH}}) = 0, 
\eq
while all the physical quantities, such as the Kretschmann scalar, Weyl scalar, $R^{ab}R_{ab}$, the energy density, and the radial and tangential pressures
of the effective energy-momentum tensor, all remain finite. Therefore, a new black hole horizon is always formed in the low curvature region. Formation of such an additional black hole horizon cannot be avoided as long as one works in a scheme where the connection components are polymerized with polymerization parameters fixed either on the entire 4-dimensional phase space (such as in \cite{Ashtekar:2005qt, Modesto:2005zm}) or on the phase space trajectories  (such as in \cite{Ashtekar:2018lag,Ashtekar:2018cay}). 
In review of the above analysis, it is clear that the existence of such a black hole horizon is tightly related to the choice of the Kantowski-Sachs gauge and the incompatibility with constant polymerization schemes when connection components are polymerized. These results demonstrate caution with generalizing these schemes to include a positive cosmological constant.  Notably the problems are of the same nature as encountered in the early quantization of LQC with a positive cosmological constant and the scheme fails to recover GR in the infra-red limit. It is important to note that the analysis provided in  Eqs.(\ref{eq4.1})-(\ref{eq4.3}) is quite general and independent of the choices of the values of the polymerization parameters as long as  $\Lambda > 0$. 

It remains to be seen whether a different polymerization approach can overcome the compatibility problems with a positive cosmological constant. One candidate can be an approach where polymerization parameters depend on phase space variables even on dynamical trajectories, such as in \cite{Boehmer:2007ket}. However, as illustrated in \cite{Ashtekar:2018lag,Ashtekar:2018cay,Saini:2016vgo,Gan:2022oiy,Zhang:2023noj},  these also face various challenges. It will be useful to further understand the interplay of gauge fixing and choices of polymerizations in this setting using insights from \cite{Giesel:2021rky} which can shed some light on compatible polymerization choice for the Kantowski-Sachs gauge. Another important issue is to check whether the assumption of validity of effective spacetime description breaks down in the presence of positive cosmological constant. Given the lack of viability of constant polymerization parameter schemes in the presence of a positive cosmological constant, and difficulties faced by other choices of polymerization parameters, our results lead to an open question whether there exists a consistent loop quantum description of Schwarzschild-de Sitter black holes.

\section*{ACKNOWLEDGMENTS}
 G.O. is supported through Baylor Physics graduate program. P.S. is supported by NSF Grants  No. PHY-2110207 and PHY-2409543, and A.W. is partly supported by the NSF Grant No. PHY-2308845.

\bibliographystyle{apsrev4-1}
\bibliography{main}

\end{document}